\documentclass[twocolumn]{aastex631}

\usepackage{color}
\usepackage{url}

\usepackage{color}
\usepackage{url}
\shorttitle{Transition state in ESO\,511-G030}
\shortauthors{Middei et al.}

\begin{document}
	
	\title{Directly tracking the re-brightening of a supermassive black hole accretion disk}
	

	
	%
	%
	%
	%
	%
	%

	
\author[0000-0001-9815-9092]{Riccardo Middei}
\correspondingauthor{Riccardo Middei}
\email{riccardo.middei@inaf.it}
\affiliation{INAF Osservatorio Astronomico di Roma, Via Frascati 33, 00078 Monte Porzio Catone (RM), Italy}

\author[0000-0001-9226-8992]{Emanuele Nardini}
\affiliation{INAF -- Osservatorio Astrofisico di Arcetri, Largo Enrico Fermi 5, I-50125 Firenze, Italy}

\author[0000-0002-1065-7239]{Chris Done}
\affiliation{Centre for Extragalactic Astronomy, Department of Physics, University of Durham, Durham, DH1 3LE, UK}

\author[ 0000-0003-0083-1157]{Elisabeta Lusso}
\affiliation{Dipartimento di Fisica e Astronomia, Universit\`a di Firenze, via G. Sansone 1, 50019 Sesto Fiorentino, Firenze, Italy}
\affiliation{INAF -- Osservatorio Astrofisico di Arcetri, Largo Enrico Fermi 5, I-50125 Firenze, Italy}

\author[0000-0002-6689-9317]{Fausto Vagnetti}
\affiliation{Dipartimento di Fisica, Universit\`a degli Studi di Roma “Tor Vergata”, via della Ricerca Scientifica 1, 00133 Roma, Italy}

\author[0000-0002-3556-977X]{Guido Risaliti}
\affiliation{Dipartimento di Fisica e Astronomia, Universit\`a di Firenze, via G. Sansone 1, 50019 Sesto Fiorentino, Firenze, Italy}
\affiliation{INAF -- Osservatorio Astrofisico di Arcetri, Largo Enrico Fermi 5, I-50125 Firenze, Italy}

\author[0000-0001-9095-2782]{Enrico Piconcelli}
\affiliation{INAF Osservatorio Astronomico di Roma, Via Frascati 33, 00078 Monte Porzio Catone (RM), Italy}

\author[0000-0002-4622-4240]{Stefano Bianchi}
\affiliation{Dipartimento di Matematica e Fisica, Universit\`a degli Studi Roma Tre, via della Vasca Navale 84, 00146 Roma, Italy}

\author[0000-0003-1994-5322]{Gabriele Matzeu}
\affiliation{Quasar Science Resources SL for ESA, (ESAC), Science Operations Department, 28692, Villanueva de la Ca\~nada, Madrid, Spain}

\author[0000-0001-8112-3464]{Anna Trindade Falc\~ao}
\affiliation{NASA Goddard Space Flight Center, Code 662, Greenbelt, MD 20771, USA}

\author[0000-0002-3626-5831]{Dominika \L. Kr\'ol}
\affiliation{Center for Astrophysics | Harvard \& Smithsonian, 60 Garden Street, Cambridge MA 02138, USA}
\affiliation{Astronomical Observatory of the Jagiellonian University, Orla 171, 30-244 Krak\'ow, Poland}

\author[0000-0003-3613-4409]{Matteo Perri}
\affiliation{INAF Osservatorio Astronomico di Roma, Via Frascati 33, 00078 Monte Porzio Catone (RM), Italy}

\author[0000-0003-3760-1910]{Alessandro Maselli}
\affiliation{INAF Osservatorio Astronomico di Roma, Via Frascati 33, 00078 Monte Porzio Catone (RM), Italy}

\author[0000-0003-1728-0304]{Keith Horne}
\affiliation{SUPA Physics and Astronomy, University of St. Andrews, Fife KY16 9SS, UK}

\author[0000-0002-6733-5556]{Juan V. Hern\'andez Santisteban}
\affiliation{SUPA Physics and Astronomy, University of St. Andrews, Fife KY16 9SS, UK}

	\begin{abstract}
		
		Accretion onto supermassive black holes powers the most luminous persistent sources in the Universe, the so-called active galactic nuclei, whose emission is characterized by two distinct spectral components: thermal optical/ultraviolet radiation from an optically thick accretion disk and a power-law X-ray tail from a corona located in the innermost regions of the accretion flow. Yet, how radiatively efficient accretion disks develop and couple to the hot corona remains poorly understood. Using six years of simultaneous ultraviolet and X-ray monitoring of the nearby active galaxy ESO\,511-G030, we witness a dramatic evolution of the broadband spectral energy distribution, driven by an increase of the ultraviolet flux from the disk by more than an order of magnitude 
		over a time scale of less than three years. The overall behavior is unlikely to track an uncovering event, and is instead compatible with a progressive recovery of the optically thick component of the accretion flow. 
		At accretion rates higher than approximately one per cent of the Eddington limit, ultraviolet and X-ray data are tightly coupled and follow the well-defined, non-linear correlation between disk and corona found in the more luminous quasars. Below this threshold, the relation apparently breaks down, as expected in case of evaporation of the inner accretion disk into a geometrically thick, optically thin hot flow. This is a strong hint of an accretion-state transition analogous to those observed in stellar-mass black holes, and confirms the need for a paradigm change in the models of radiatively efficient accretion flows around supermassive black holes.
		
	\end{abstract}
	

	\section{Introduction} \label{sec:intro}

	Accretion onto compact objects is a fundamental process powering a wide range of high-energy astrophysical sources, from X-ray binaries (XRBs), where matter from an evolved companion star is drawn towards a stellar-mass black hole or neutron star, to the luminous cores of Seyfert galaxies and quasars, collectively known as active galactic nuclei (AGN). In the latter systems, the release of gravitational potential energy through accretion onto a supermassive black hole (SMBH) produces copious radiation, mainly emitted in the optical/UV band by the accretion disk \citep{Shields1978} and in X-rays by the so-called hot corona, a $\approx$\,10$^9$ K plasma that intercepts and inverse-Compton scatters the disk photons \citep{Haardt1991,Haardt1993}.
	
	Both the luminosity and the spectral energy distribution (SED) of this emission depend on the mass of the black hole ($M_{\rm BH}$) and on its accretion rate relative to the Eddington limit ($\lambda_{\mathrm{Edd}}=L/L_{\mathrm{Edd}}$, where $L$ is the total source luminosity and $L_{\mathrm{Edd}}$\,$\simeq$\,1.26\,$\times$\,10$^{38}$\,$M_{\rm BH}/M_{\odot}$), i.e., the physical limit at which gravity is balanced by radiation pressure. At super-Eddington accretion rates ($\lambda_{\mathrm{Edd}}$\,$>$\,1), radiation pressure dominates the inner regions of the flow, leading to geometrically and optically thick disks that can drive powerful  winds. This form of mechanical energy extraction, together with photon trapping and advection phenomena, greatly reduces the radiative efficiency, likely setting a cap to the source luminosity \citep[e.g.,][]{Abramowicz1988, Wang1999, Poutanen2007}. 
	At intermediate accretion rates (0.01\,$\lesssim$\,$\lambda_{\mathrm{Edd}}$\,$\lesssim$\,1), the accretion flow is instead expected to form a geometrically thin and optically thick disk \citep[][hereafter SS73]{Shakura1973}, which constitutes the most efficient way of radiating away the gravitational energy of the infalling matter, and results in a prominent thermal emission component peaking at temperatures $T$\,$\propto$\,$ (\lambda_{\mathrm{Edd}} /M_{\rm BH})^{1/4}$. 
	At low accretion rates ($\lambda_{\mathrm{Edd}}$\,$\lesssim$\,0.01), the radiative efficiency drops again, following the transition to a geometrically thick, optically thin mode such as an advection-dominated accretion flow \citep[ADAF;][]{Narayan1995}. 
	
	These different accretion regimes can be clearly seen in stellar-mass BH systems, whose SEDs are dominated by an optically thick disk emission, peaking at 2--3~keV when the source is in a bright state ($\lambda_{\mathrm{Edd}}$\,$\sim$\,0.5) and cooling as predicted by disk models as the luminosity drops to $\lambda_{\mathrm{Edd}}$\,$\sim$\,0.01. At even lower accretion rate values, the emission from the inner disk almost completely vanishes, in agreement with a scenario where the optically thick, geometrically thin disk is replaced by a radiatively inefficient ADAF \citep{Done2007}. Analogous accretion states in SMBHs have been tentatively identified only in population studies so far \citep[e.g.,][]{Svoboda2017,Ruan2019}. Remarkably, X-ray unobscured AGN with $\lambda_{\mathrm{Edd}}$\,$\gtrsim$\,0.01 display an optical/UV spectrum rising at shorter wavelengths as predicted by SS73-like disk models, while this feature disappears below $\lambda_{\mathrm{Edd}}$\,$\lesssim$\,0.01 \citep[][]{Hagen2024,Kang2025}.
	
	However, the transition itself is extremely difficult to monitor in real time in an individual AGN. This is mainly due to two reasons. First, the transition is intrinsically hard to identify in SMBHs, as the inner-disk temperature typically peaks at extreme UV energies ($\approx$\,1000 \AA, or 0.01 keV; \citealt{CaiWang2023}). Second, variability is predicted to obey the viscous time scale, which linearly grows with $M_{\rm BH}$ at fixed radial distance. 
	While stellar-mass BHs undergo this transition on time scales of less than one day, the same process is expected to take thousands of years or longer in SMBHs with $M_{\rm BH}\sim10^{6-8} M_{\odot}$ \citep{Noda2018}.
	
	Here we present the results of a six-year monitoring campaign of the nearby ($z$\,=\,0.023; \citealt{Koss2022}) active galaxy ESO\,511-G030, during which we witness an increase of the UV flux from the nuclear component by about a factor of $\approx$\,30, accompanied by a similar, long-term rise by an order of magnitude of the X-ray flux. We interpret this exceptional behavior as due to the real-time recovery of an optically thick phase in the accretion flow following a drop in the mass supply through the outer accretion disk. 
	
	\section{Swift monitoring of ESO\,511-G030}
	ESO\,511-G030 is a type-1 Seyfert nucleus hosted by a nearly face-on spiral galaxy \citep[][]{deVaucouleurs1991}, in which a black-hole mass of $M_{\rm BH}$\,$\sim$\,1.7\,$\times$\,10$^{7}$\,$M_{\odot}$ was estimated from the width of the broad H$\beta$ component \citep{Koss2022}. 
	ESO\,511-G030 is known to be a `bare' AGN, in which no X-ray absorption in excess of the Galactic value has been found \citep[e.g.,][]{Laha2014}. Together with the favorable orientation, this suggests that also intrinsic dust reddening, if any, should be minimal. Indeed, a past {\it XMM-Newton} observation carried out in 2007 caught ESO\,511-G030 in a bright state in both the UV and X-rays. Conversely, when the source was observed again by {\it XMM-Newton} in 2019, the AGN was found in an X-ray flux state approximately ten times fainter, with a comparable fading also observed in the UV emission \citep[][]{Middei2023}.

	Starting from 2019, we thus embarked on a systematic monitoring of this source with the Neil Gehrels Swift Observatory \citep[hereafter {\it Swift};][]{Gehrels2004}, consisting of more than 80 observations between August 2019 and September 2025, for a total net exposure of $\sim$\,390 ks.\footnote{Individual exposures span $\sim$\,0.8--9.8 ks, with a typical duration of $\sim$\,4--5 ks.} The monitoring strategy corresponds to a mean cadence of roughly one visit per month occasionally increasing to a bi-weekly sampling, excluding the seasonal visibility gap that occurs October--November. This dense and rather uniform {\it Swift} campaign has provided simultaneous optical/UV (accretion disk) and X-ray (corona) measurements of ESO\,511-G030, enabling us to track with unprecedented detail the long-term spectral evolution of this AGN, possibly characterized by a dramatic change in its accretion state.

	\section{Data reduction}
	\begin{figure*}[t]
		\centering
		\includegraphics[width=0.99\textwidth]{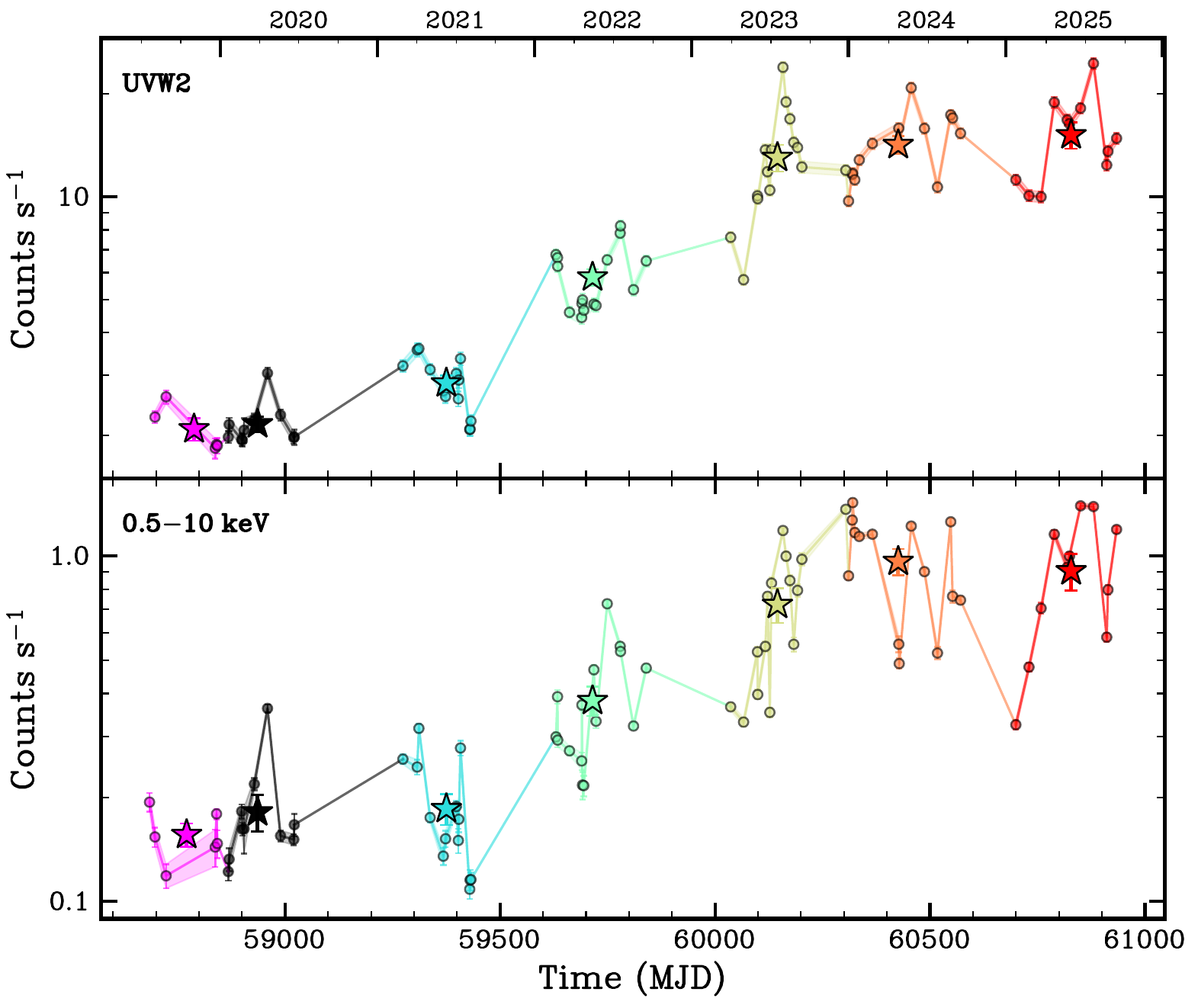}
		\caption{UV (W2; effective $\lambda$\,=\,1991 \AA) and X-ray (2--10 keV) time evolution. \textit{Top:} observed count-rate light curve for the UVW2 filter. \textit{Bottom:} same for the 0.5--10 keV X-rays. Stars account for the mean flux within the annual time bin. The colors indicate the calendar year of observation; the same code is used in all the following plots.}
		\label{figlcs}
	\end{figure*}
	
	The uniquely rich data set from the 2019--2025 \textit{Swift} monitoring of ESO\,511-G030 consists of 82 epochs. The source was simultaneously observed with the X-ray Telescope \citep[XRT;][]{Burrows2005} in photon counting mode, and with the UltraViolet/Optical Telescope  \citep[UVOT;][]{Roming2005}, for which all the filters were requested.\footnote{Although the campaign comprises a total of 84 observations, the first two exposures obtained in 2019 were excluded from the analysis because not all optical/UV filters were available.}
	
	The high-level science products for each \textit{Swift}-XRT observation were generated using an automated pipeline that retrieves and processes the raw event data. This workflow relies on the standard \textit{Swift}-XRT analysis tasks, namely \textsc{xrtpipeline} and \textsc{xrtproducts}, as described in the reference guide by Capalbi et al.~(2005).\footnote{\url{https://xep-swiftv-web.star.le.ac.uk/analysis/xrt/files/xrt_swguide_v1_2.pdf}} 
	The source and background extraction regions used for spectral and timing analyses were visually inspected for all observations. Source counts were extracted from a circular region with a fixed radius of 30 pixels (corresponding to $\sim$\,70 arcsec). The background was extracted from an annular region centered on the source position, whose inner radius was offset by 25 pixels ($\sim$\,60 arcsec) from the source region. Finally, the spectra were grouped to ensure a minimum of 5 counts per bin.
	
	To extract the count rates from each UVOT exposure, we used a similar automated procedure. For each available filter, we summed all exposures using the command \textsc{uvotimsum}. 
	Then we relied on the \textsc{uvot2pha} procedure to extract the count rates for the source and the background. In particular, two circular regions were used: the first centered on the source (with radius of 4 arcsec) and the second one (with radius of 15 arcsec) covering a source-free region of the detector. The rates obtained in this way for the UVOT filters are the observed ones, i.e., not corrected for Galactic reddening. 
	
	In Fig.~\ref{figlcs} we show the evolution of UV and X-ray count rates over the entire campaign. Both the UV and X-ray light curves are characterized by an overall monotonic flux increase by about a factor of $\sim$\,10 (note that the UV emission of the host galaxy is not subtracted at this stage). While small-amplitude flickering on shorter time scales (weeks/months) is also observed, the long-term trend most likely traces an evolution in the accretion state, as we argue in the following.
	
	\section{The origin of the UV brightening}
	
\begin{figure*}[h]
		\centering
		\includegraphics[width=0.99\textwidth]{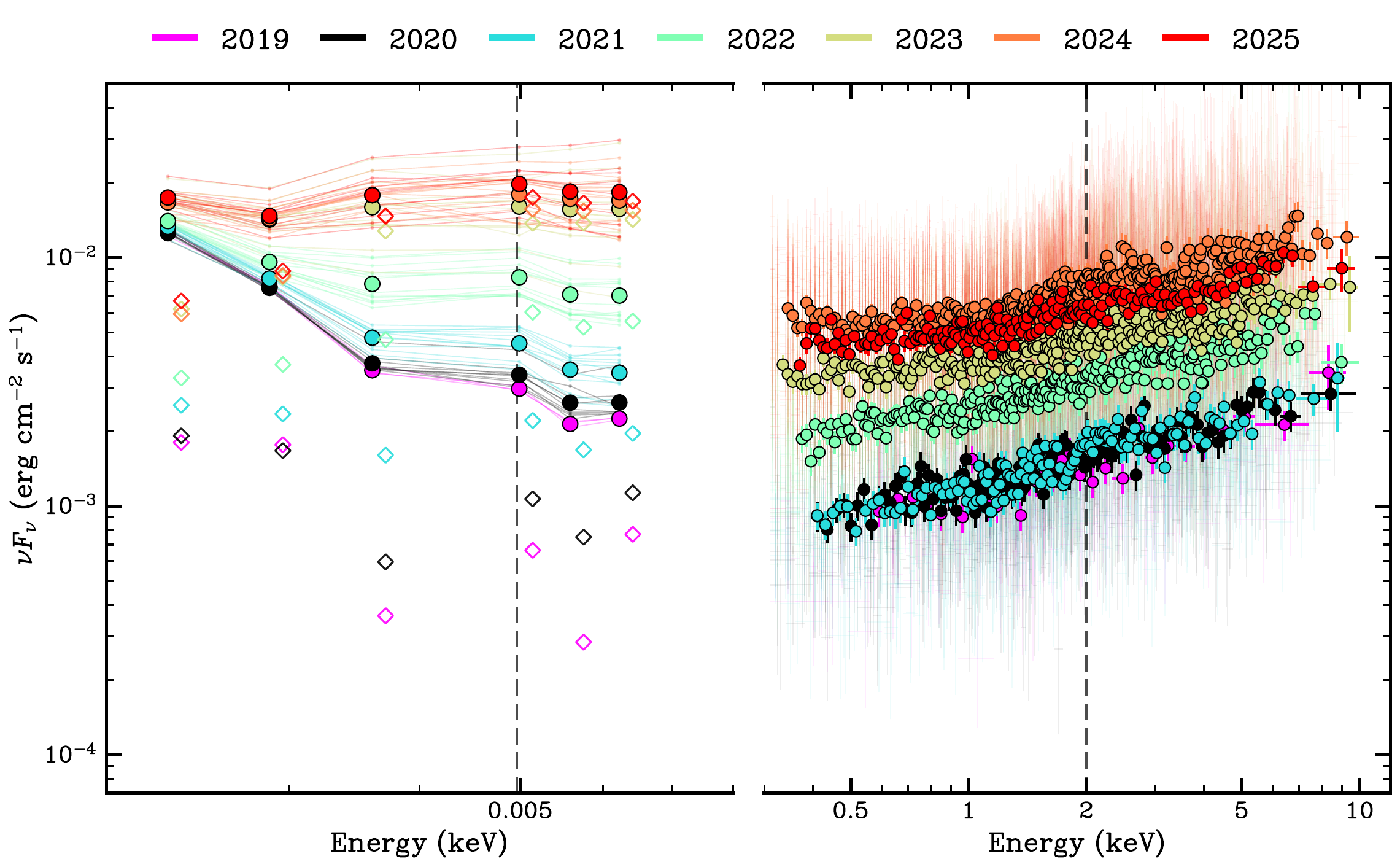}
		\caption{Change in the observed broadband spectral energy distribution of ESO\,511-G030 during the {\it Swift} monitoring. Both the UVOT (left panel) and XRT (right panel) spectra are intrinsic, i.e., corrected for extinction and absorption based on the literature values of Galactic reddening and hydrogen column density. The thin background symbols represent all the individual SEDs of the source. The data are also binned on a yearly time scale (dots) to better highlight the long-term evolution, with the averaged X-ray spectra further grouped adaptively based on the signal-to-noise for visual purposes. The empty diamonds refer to the host-galaxy subtracted flux levels (see Sec.~\ref{galaxy} for details; the symbols are slightly shifted in energy for the sake of clarity). The effective wavelengths of the UVOT filters are the following: 5410 \AA\ (V), 4321 \AA\ (B), 3442 \AA\ (U), 2486 \AA\ (W1), 2221 \AA\ (M2), 1991 \AA\ (W2), after   \citet{Breeveld2011}.} 
		\label{fig:vfv}
	\end{figure*}

	These multi-epoch, multi-frequency {\it Swift} observations demonstrate that the broadband flux rise is a manifestation of the conspicuous evolution with time of the shape of the optical-to-X-ray SED of ESO\,511-G030, as illustrated in Fig.~\ref{fig:vfv}. 
	In this plot, the light background shades account for each single observation obtained during the monitoring campaign. To average out the short-term fluctuations and simultaneously reduce the arbitrariness in the definition of the time intervals, we considered each calendar year from 2019 to 2025 to define seven broadband spectra. Here the data are also corrected for both Galactic reddening \citep{Schlafly2011} and interstellar absorption \citep{HI4PI2016}. The UVOT count rates were converted into monochromatic fluxes at the effective wavelengths of each filter following the conversion factors reported in the official calibrations documents.\footnote{\url{https://heasarc.gsfc.nasa.gov/docs/heasarc/caldb/swift/docs/uvot/uvot_caldb_AB_10wa.pdf}}
	
	Overall, the reddest UVOT filter (V; central $\lambda$\,=\,5410 \AA) undergoes a $\sim$\,35\% flux rise in the yearly averages over the course of the campaign, while for the bluest one (UVW2; effective $\lambda$\,=\,1991 \AA) the increase is by almost a factor of eight (or 2.2 mag). Conservatively assuming that the host-galaxy contamination within the latter filter is negligible, this sets a hard lower limit to the enhancement of the UV emission from the AGN. More specifically, at early times (magenta and black) the optical/UV emission is possibly dominated by the host galaxy in every filter, while the X-ray emission is well described by a single power-law component with a hard ($\Gamma$\,$\sim$\,1.6) photon index. The UV luminosity then starts to increase (cyan) without being immediately followed by the X-ray one (the cyan, magenta, and black X-ray spectra lie on top of each other). 
	The rapid increase in the UV flux in 2022--2023 (mint, gold) is closely matched in the X-rays instead. During this period, the source experienced the most remarkable change in the overall SED, driven by the contribution from the AGN disk that largely outshines any emission from the host galaxy in the UV. Since 2023 (orange and red), the X-ray spectra begin to show a progressive softening below 2 keV. While the eventual development of a soft excess strongly supports an almost complete recovery of the inner disk as the accretion rate grows \citep[e.g.,][]{Jana2026}, other effects might also be involved.

	\subsection{Progressive nuclear uncovering}
	\label{uncover}
	
\begin{figure*}[t]
		\centering
		\includegraphics[width=0.99\columnwidth]{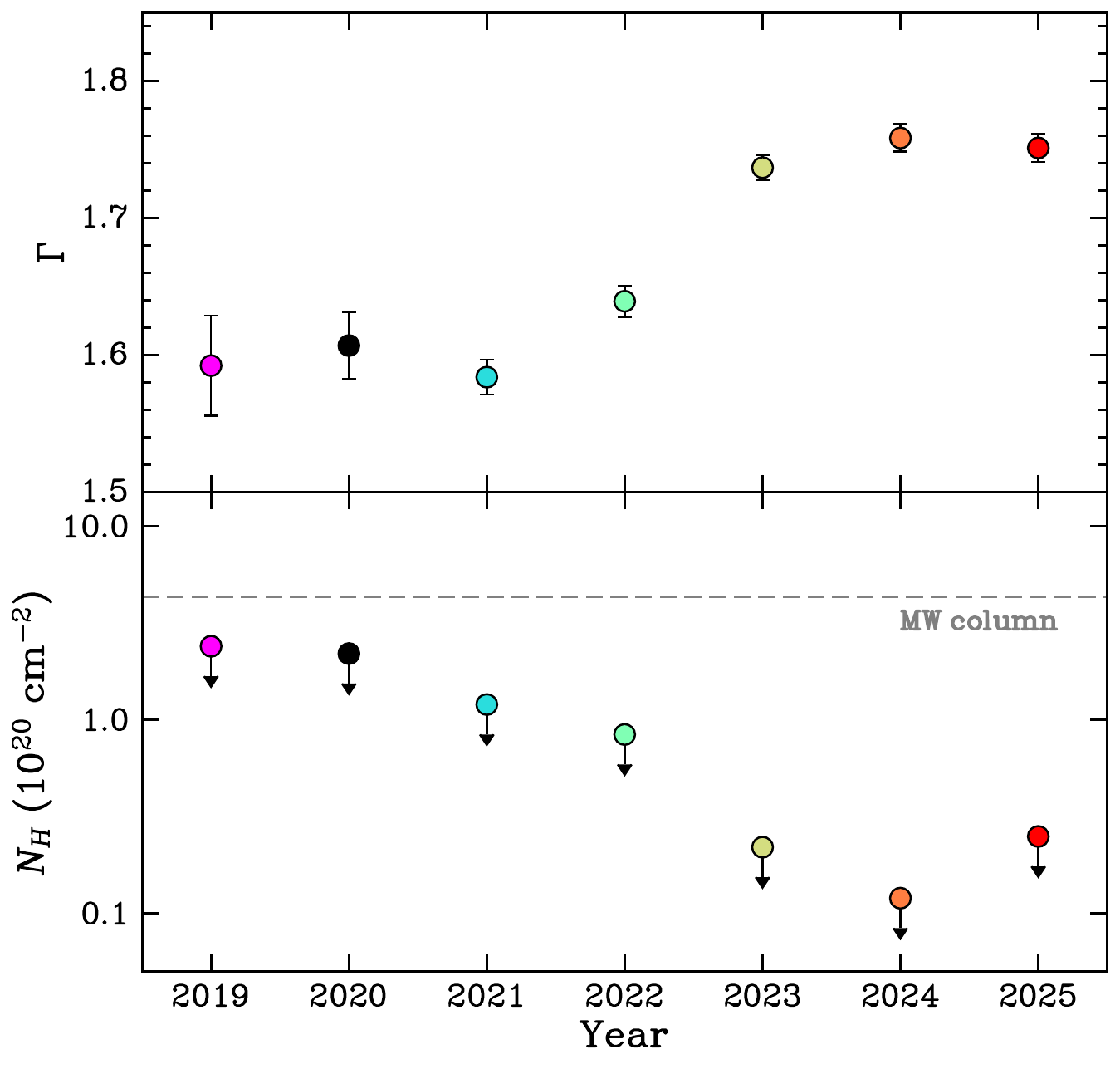}
		\includegraphics[width=0.99\columnwidth]{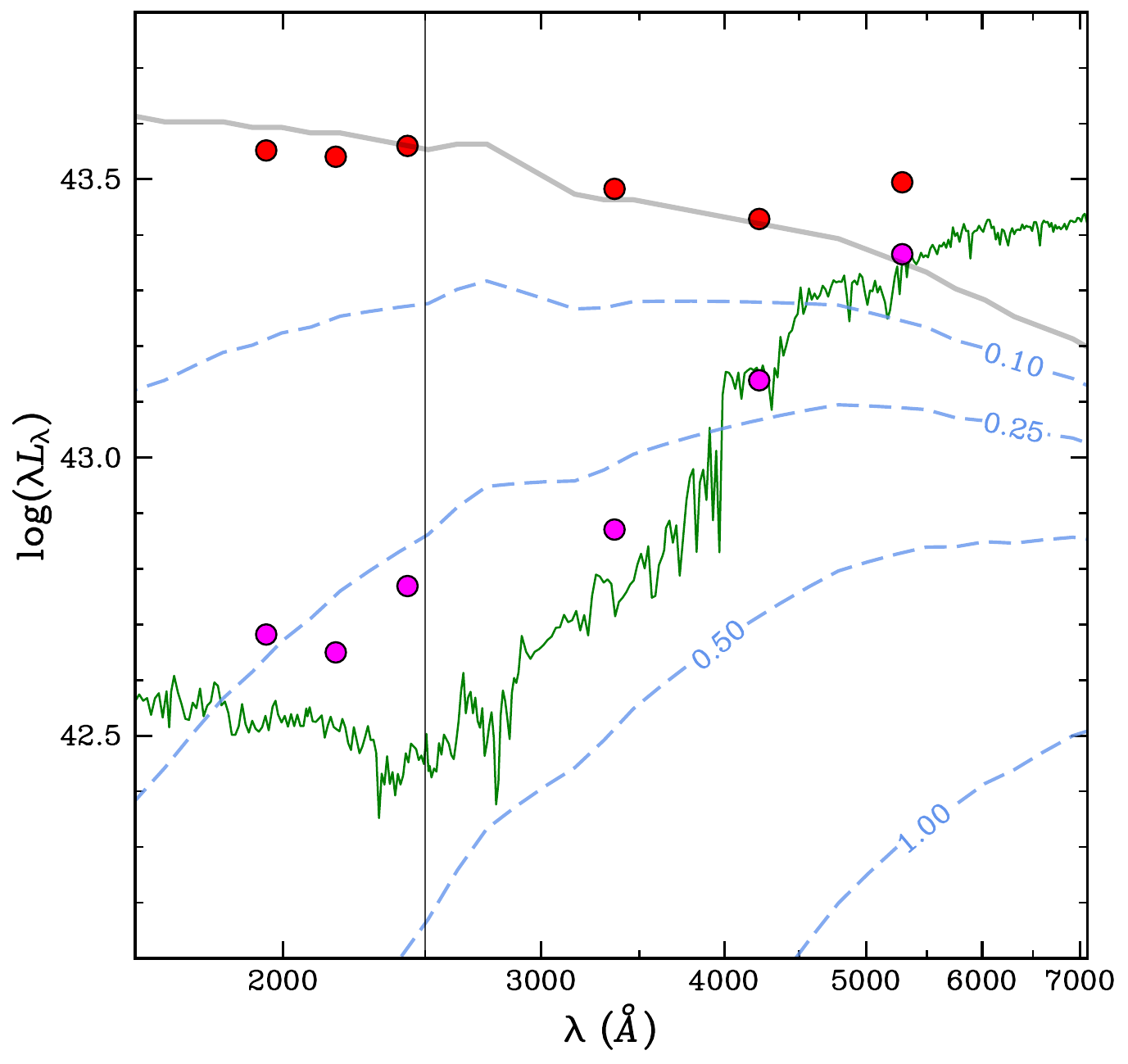}
		\caption{{\it Left panel}: Evolution of the X-ray (0.5--10 keV) photon index $\Gamma$ from the average \textit{Swift}-XRT spectra (top) and 3$\sigma$ upper limits on any hydrogen column density local to the source $N_{\rm H}(z)$ in excess of the Galactic value (bottom). Across the whole campaign, the X-ray spectra imply minimal optical/UV extinction for a standard Galactic gas-to-dust ratio. {\it Right panel}: Optical/UV SED of ESO\,511-G030 from \textit{Swift}-UVOT in the low (2019; magenta dots) and high state (2025; red dots). The gray solid line is the intrinsic AGN template from \citealt{Saccheo2023}, that best describe the high-state data, while the light-blue dashed curves show the effects of increasing reddening up to $E(B-V)$\,=\,1 assuming an SMC extinction law. While no reddened AGN SED is acceptable, the low-state optical data are consistent with being dominated by the host galaxy (green spectrum; see Section \ref{galaxy} for details).}
		\label{fig:obscure}
		\label{fig:sed_reddening}
	\end{figure*}

	In principle, variable obscuration represents another mechanisms capable of producing significant changes in the UV flux over time scales of several years. We thus explore the possibility that the low-flux states observed during 2019--2021 are due to an intervening screen of dust and gas along the line of sight, which then progressively unveils the inner regions of the accretion flow. Even without considering the problematic geometrical and dynamical aspects, we can safely rule out the latter scenario for ESO\,511-G030. We start by noting that any foreground obscurer that covers a substantial fraction of the UV-emitting region of the disk will almost inevitably hide also the much more compact X-ray corona. Yet, even in the low state, the X-ray spectra have a simple power-law shape with no evidence for a low-energy cut-off due to absorption from cold gas. We quantified this attribute by analyzing the \textit{Swift}-XRT spectra within the spectral fitting package \textsc{Xspec} \citep[][]{Arnaud1996}. We modeled the averaged 0.5--10 keV X-ray spectra with a power law, $F(E)\propto E^{-\Gamma}$, where we allowed for an absorbing column density local to the source in excess of the Galactic one ($N_{\rm H}$\,=\,4.3\,$\times$\,10$^{20}$ cm$^{-2}$; \citealt{HI4PI2016}). While the continuum becomes progressively softer with time (consistent with a rise of the accretion rate; e.g., \citealt{Trakhtenbrot2017}), we could only place 3$\sigma$ upper limits on any intrinsic column density $N_{\rm H}(z)$, the more stringent the better the spectral quality (Fig.~\ref{fig:obscure}, left panel). Converting the limits on $N_{\rm H}(z)$ to reddening assuming a Galactic dust-to-gas ratio \citep{Bohlin1978}, we obtain $E(B-V)$\,$\sim$\,1.7\,$\times$\,10$^{-22}\,N_{\rm H}$\,$\lesssim$\,0.04 (at 3$\sigma$) in 2019. It is worth reminding that the $E(B-V)/N_{\rm H}$ ratio in AGN is generally much lower than the standard Galactic value \citep{Maiolino2001}, hence even a modest extinction in the optical/UV should be accompanied by an easily detectable photo-electric cut-off in the X-rays.
	
	Any configuration where the disk emission is attenuated while the coronal one is not would be highly contrived. Even so, the observed UVOT trend is in itself incompatible with an unveiling of the disk. To demonstrate this, we analyzed the optical/UV SED shape in the extreme 2019 and 2025 states, under the hypothesis that the SED evolution can be solely ascribed to dust. We compared the high state with the type-1 AGN template from \citet{Saccheo2023} that best matches the short-wavelength data points (Fig.~\ref{fig:obscure}, right panel), which can be reasonably assumed to be dominated by the nuclear component in 2023--2025.\footnote{We note that the V-band flux excess seen in the bright phase compared to the \citet{Saccheo2023} template is mostly due to the larger host-galaxy contribution, with some contamination from the [O~\textsc{iii}] doublet and the blue wing of the broad H$\beta$ line.} We then applied a Small Magellanic Cloud (SMC) extinction law \citep{Prevot1984}, as appropriate for AGN \citep[e.g.,][]{Hopkins2004}, with color excess values ranging from 0 to 1 in steps of 0.05. Some representative reddened AGN SEDs are shown in (Fig.~\ref{fig:obscure}), from which it clearly emerges that the low state of ESO\,511-G030 cannot be reproduced in a simple scenario of variable AGN obscuration. Interestingly, a galaxy template (see below) would adequately account for the low-state optical data instead. As no combination of galaxy plus reddened AGN can describe the entire UVOT SED, this suggests that the nuclear output might have indeed dropped below the stellar emission intrinsically. 
	
	\subsection{Host-galaxy contamination}
	\label{galaxy}
	
\begin{figure*}[t]
		\centering
		\includegraphics[width=0.99\textwidth]{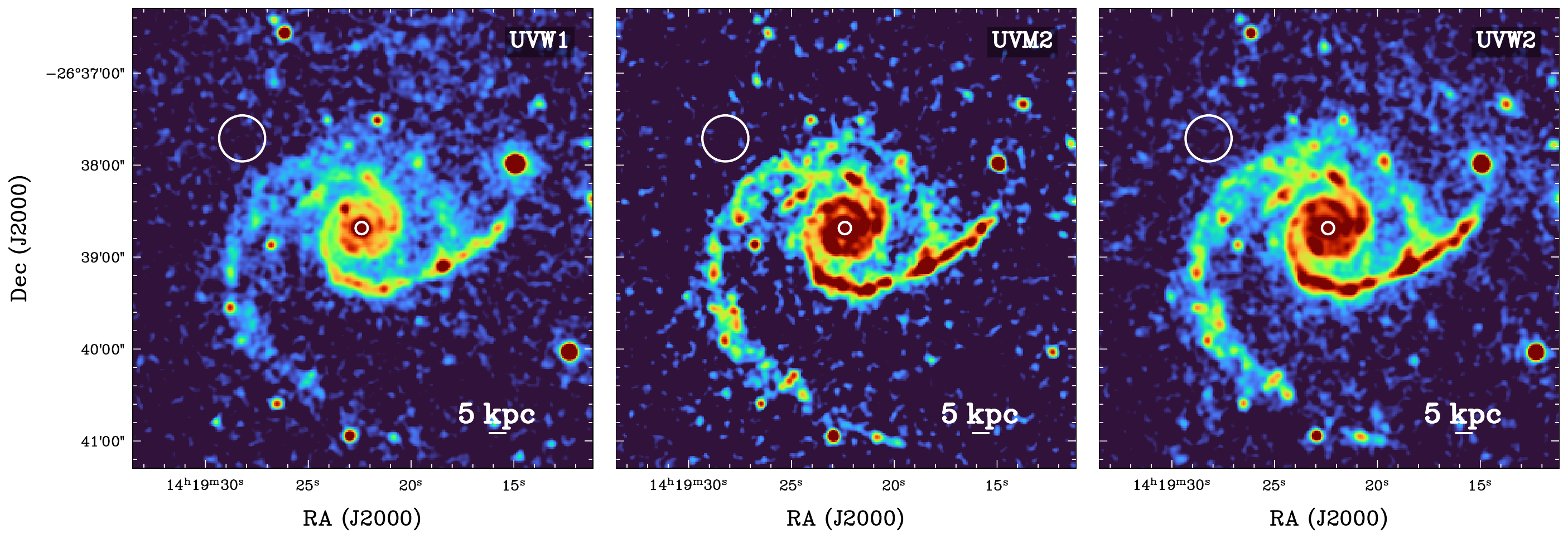}
		\caption{Stacked multi-filter (UVW1, UVM2, UVW2) images of ESO\,511-G030 taken in 2019. Prominent spiral arms and star formation knots are clearly identified, consistent with the late-type morphological classification of the host galaxy \citep[Sc;][]{Lauberts1982}. The images are aligned to a common WCS reference frame and smoothed with a Gaussian kernel ($\sigma$\,=\,1.5 pixels) to enhance the S/N ratio of the extended features. The scale bar corresponds to a physical distance of 5~kpc at the redshift of ESO\,511-G030 ($z$\,=\,0.0223), while the circles indicate the source and background extraction regions, respectively.}
		\label{fig:uvotsc}
	\end{figure*}

	Having established that the behavior we are witnessing between 2019 and 2025 can be confidently interpreted as a resurgence of the AGN disk emission, we need to assess the (constant) contribution of the host galaxy to the UVOT photometry in order to determine the actual span of the AGN variability. To this aim, we employed low-resolution observations of ESO\,511-G030 obtained between 2019 and 2025 with the FLOYDS spectrographs on the 2-m Faulkes Telescope North at Haleakala, Hawaii (USA), and the identical 2-m Faulkes Telescope South at Siding Spring (Australia), part of the Las Cumbres Observatory network. We provide the basic details on the acquisition and reduction of the latter data in Appendix~\ref{floyds}, noting that the analysis of the spectral evolution of the source in the optical goes beyond the scope of the present work, and will be presented in a subsequent paper. In short, we fitted all the available spectra separately, using the spectral decomposition code \textssc{QSFit} \citep{Calderone2017} and adopting a SWIRE Sc galaxy template \citep{Polletta2007}. This choice follows the classification of ESO\,511-G030 \citep{Lauberts1982} and is supported by the visual inspection of the host galaxy, which reveals a late-type, spiral morphology characterized by prominent, open spiral arms and numerous sites of intense star formation (Fig.~\ref{fig:uvotsc}). Such regions significantly contribute to the UV flux, especially during the AGN low state, making the Sc template particularly suitable for an accurate host-galaxy subtraction.
	
	The AGN continuum was modeled as a power law of the form $F_{\lambda}$\,$\propto$\,$\lambda^{-1.7}$, where the slope was fixed to a canonical value. Indeed, as discussed by \citet{Calderone2017}, the exact shape of the AGN continuum cannot be better constrained over a limited bandwidth in the presence of a substantial contribution from the host galaxy (see Fig.~\ref{fig:sed_reddening}). The key parameter extracted from the fit was therefore the relative weight of the power-law continuum and host-galaxy template. The inferred host-galaxy level, averaged and extrapolated to rest-frame 2500 \AA, is $L_{\rm UV,gal}$\,$\sim$\,2.5\,$\times$\,10$^{27}$ erg s$^{-1}$ Hz$^{-1}$ (see Appendix~\ref{floyds}), implying that the galaxy accounts for up to $\approx$\,70\% of the observed UV flux in the low state. To further validate this estimate, we analyzed the radial surface brightness of the stacked 2019 UVOT images in each UV filter (Fig.~\ref{fig:uvotsc}), and fitted the profile with two Gaussian components: a narrow one representing the AGN in the core, and a broad one describing the host-galaxy bulge. We found that the the extended component alone accounts for $\sim$\,15--20\% of the total flux within the central region of 4-arcsec radius (i.e., almost 2 kpc at the distance of ESO\,511-G030). We stress that this fraction should be considered as a lower limit, since Sc galaxies usually exhibit a prominent central cusp in their surface brightness profiles, which cannot be spatially resolved here (see \citealt{Bouquin2018}, based on {\it GALEX} data). This suggests that the flux encompassed by the compact Gaussian component can still be highly contaminated by the the innermost stellar populations of the host, confirming that our estimate of the galactic contribution $L_{\rm UV,gal}$ as obtained from the extrapolation of the FLOYDS spectra is fully credible within the measurement uncertainties.
	
	\subsection{Accretion state transition}
	\label{csagn}
	
	As the host galaxy is responsible for a non-negligible, and possibly major share of the observed emission in the low state, the actual brightening of the AGN disk component in the UV likely amounts to a factor of $\sim$\,20--30 (cf.~with the empty diamonds in the left panel of Fig.~\ref{fig:vfv}). The mutual behavior of the UV and X-ray emission can provide useful clues on the properties of the accretion flow during this exceptional evolution. 
	A well established way to study the interplay between the accretion disk and the hot corona in AGN is represented by the $L_{\rm X}$--$\log L_{\rm UV}$ relation \citep[e.g.,][]{Lusso2016}. The UV and X-ray bright AGN, which are believed to accrete efficiently via a fully formed SS73-like disk, abide by a tight correlation between the X-ray coronal flux at 2~keV and the optically thick disk emission at 2500 \AA\ (indicated by the rightmost and leftmost vertical dashed lines in Fig.~\ref{fig:vfv}, respectively), whereby $L_{\rm X}$\,$\propto$\,$L_{\rm UV}^{\gamma}$ with $\gamma$\,$\sim$\,0.6. 
	
	The individual \textit{Swift}-XRT spectra from each visit were used to estimate the luminosity at 2 keV, which is always dominated by the AGN even in the low state. Following the previous analysis, which rules out X-ray absorption in the source frame, we adopted a simple power-law model with Galactic column, in which the continuum photon index and normalization were left free to vary. Such a model yields a satisfactory description ($C_{\rm stat}$/d.o.f.\,$\sim$\,1) of all the spectra, hence we applied the \texttt{clumin\footnote{\url{https://heasarc.gsfc.nasa.gov/docs/software/xspec/manual/node305.html}}} component to the best fits to derive the intrinsic continuum luminosity, which was subsequently converted into a monochromatic value at 2 keV, $L_{\rm X}$, with its associated uncertainty.
	
	The equivalent quantity at 2500 \AA\ was extracted from the interpolation of the linear expression $\log{L_\nu}$\,=\,$a\log{\nu}+b$, fitted to the six UVOT photometric points at each epoch. The associated uncertainty was evaluated through a Monte Carlo approach, starting from the observed data and generating 2000 synthetic realizations of each SED by perturbing every $[\log{\nu},\log L_{\nu}]$ point with random noise consistent with the observed dispersion. The simulated SEDs were then re-fitted with the same model to build the distributions of the parameters $a$ and $b$ and, consequently, of the interpolated luminosity at 2500 \AA, the uncertainty on which incorporates the effects of both the measurement noise and the natural scatter of the SED shape. We finally subtracted the host-galaxy contribution calculated in Sec.~\ref{galaxy} and obtained the intrinsic emission from the AGN at 2500~\AA, $L_{\rm UV}$.
	
\begin{figure*}[h]
		\centering
		\includegraphics[width=0.99\textwidth]{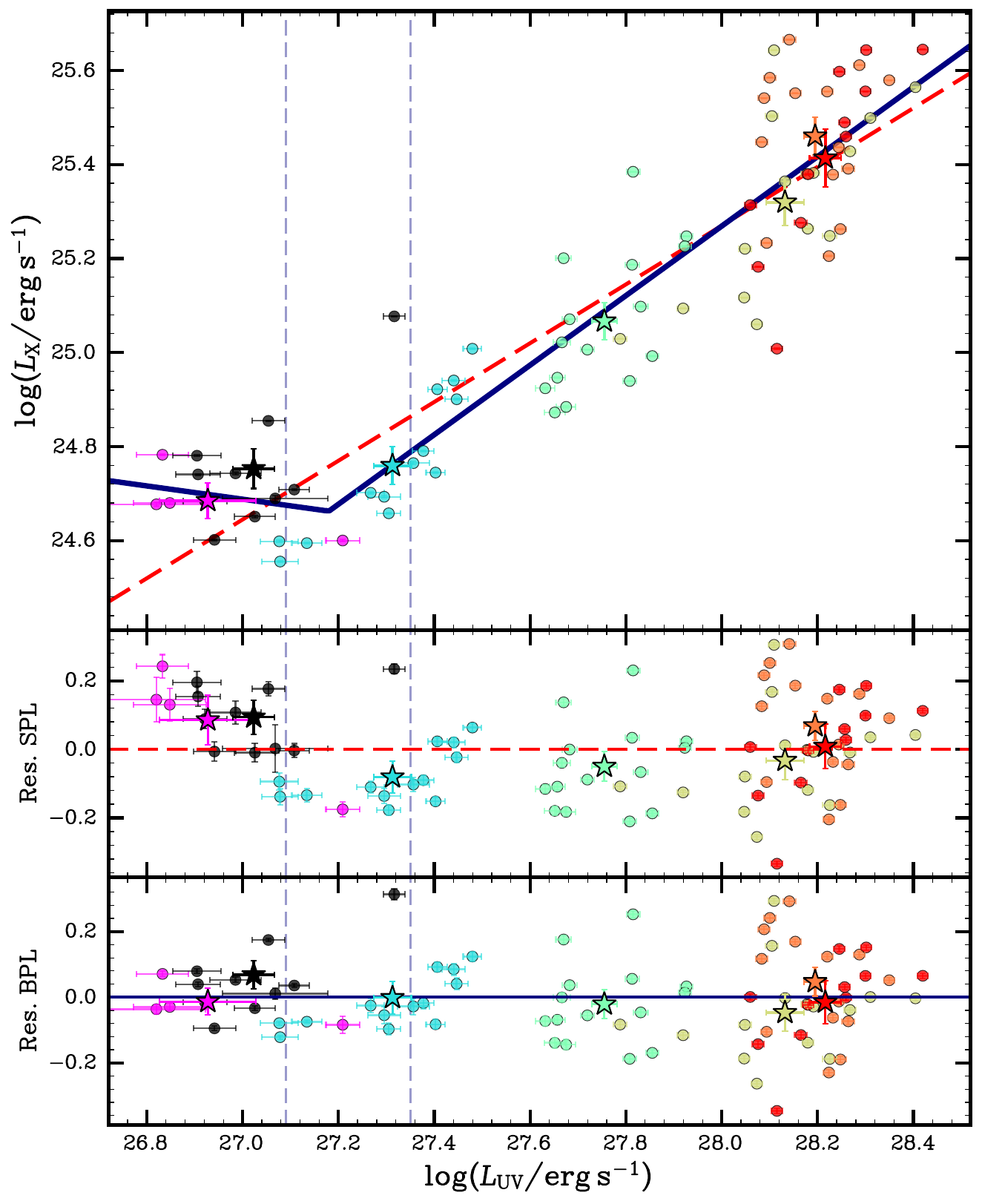}
		\caption{$L_{\rm X}$--$L_{\rm UV}$ relation obtained from our data of ESO\,511-G030, where both quantities are AGN monochromatic luminosities, at 2 keV and 2500 \AA, respectively. Two distinct accretion regimes can be tentatively identified: one consistent with the standard $L_{\rm X}$--$L_{\rm UV}$ relation found for the general AGN population \citep[][]{Lusso2020}, and another that significantly deviates from the latter at low UV luminosities, where the X-ray flux remains roughly constant despite the decrease in UV flux. This behavior hints at a departure from the standard disc--corona energy balance, suggesting the transition to a different accretion regime. Vertical dashed lines are used to mark the uncertainties of the $L_{\rm UV,break}$.}
		\label{fig:LxLuv}
	\end{figure*}

	Fig.~\ref{fig:LxLuv} shows the $\log L_{\rm X}$--$\log L_{\rm UV}$ relation for ESO\,511-G030 based on our {\it Swift} observations. At a preliminary visual inspection, the data points apparently deviate from a simple non-linear slope when the disk luminosity in ESO\,511-G030 falls below $L_{\rm UV}$\,$\approx$\,10$^{27.3}$ erg s$^{-1}$ Hz$^{-1}$, where the X-ray output decouples from the decreasing UV trend. 
	This coincides with the initial delay in the recovery of the X-ray emission mentioned above, and suggests a transition in the accretion flow occurring around the end of 2021 (see also Fig.~\ref{figlcs}), when the connection between the disk emission and the coronal one appears to change. 
	
	Following \citet{Lusso2016}, who described the $L_{\rm X}$--$L_{\rm UV}$ correlation in a statistically significant sample of quasars through a linear fit in the log--log plane with a slope $\gamma$\,=\,0.642$\pm$0.015 and an intrinsic dispersion $\delta$\,=\,0.24 dex, we investigated the time evolution of the $L_{\rm X}$--$L_{\rm UV}$ relation in ESO\,511-G030 performing a Bayesian regression analysis. We first tested the standard single power-law (SPL) model: $\log L_{\rm X}$\,=\,$\gamma \log L_{\rm UV}$\,+\,$\beta$. The parameters were estimated by employing the \texttt{emcee} Markov Chain Monte Carlo (MCMC) sampler \citep{Foreman-Mackey2013}, adopting a likelihood function that self-consistently accounts for the uncertainties in both variables and includes the intrinsic scatter ($\delta_{\rm int}$) as a free parameter. We used 64 walkers for 2500 steps, discarding the first 700 steps as burn-in. The final parameters and their $1\sigma$ uncertainties were derived from the 16$^{\rm th}$, 50$^{\rm th}$, and 84$^{\rm th}$ percentiles of the posterior distributions. The inferred slope for this linear test, $\gamma$\,=\,0.62$\pm$0.03, is fully consistent with that observed in the general population of blue quasars (see also \citealt{Lusso2020}). However, as shown by the residuals in the central panel of Fig.~\ref{fig:LxLuv}, this fit leaves a clear structure of poorly modeled data in the low-$L_{\rm UV}$ regime. 
	
	To better characterize the overall behavior of ESO\,511-G030, we thus explored a more flexible parameterization of the $L_{\rm X}$--$L_{\rm UV}$ relation by introducing a second-order polynomial function, with three free coefficients. 
	By adopting the same MCMC procedure, we find that the polynomial model provides a better description of the data, which, according to a likelihood-ratio test \citep{Wilks1938}, corresponds to a statistically significant departure from a simple linear scaling in log--log space, with a $p$-value of $\approx$\,2\,$\times$\,10$^{-4}$. Motivated by this hint of non-linearity, and aiming at a more physically informative parameterization, we then tested a broken power-law (BPL) model defined by two slopes, $\gamma_1$ and $\gamma_2$, separated by a break luminosity $L_{\rm UV,break}$. Also the BPL model results in a definite improvement over the SPL one (see bottom panel of Fig.~\ref{fig:LxLuv}), 
	allowing us to reject the null hypothesis with high statistical significance ($p$-value of $\approx$\,3\,$\times$\,10$^{-4}$, or $\sim$\,3.6$\sigma$). This preference is further supported by the Bayesian Information Criterion \citep[][]{Kass1995}, with a $\Delta$BIC\,=\,7.44 (only marginally worse than the polynomial case, for which $\Delta$BIC\,$\sim$\,9.3), again indicating the limitations of the linear model.
	
	While both the polynomial and BPL models acceptably describe the time-resolved $L_{\rm X}$--$L_{\rm UV}$ relation in ESO\,511-G030, the latter represents a directly interpretable framework, as it naturally distinguishes between two distinct regimes. The best-fit BPL model parameters are $\gamma_1$\,=\,$ -$0.15$\pm$0.30 below, and $\gamma_2$\,=\,0.74$\pm$0.07 above the break luminosity of $\log L_{\rm UV,break}$\,=\,27.18$^{+0.17}_{-0.09}$. This turning point in the UV luminosity of ESO\,511-G030 is reached during 2021, while from 2022 onward the UV luminosity has consistently remained well above that threshold.

	\section{Discussion and Conclusions}\label{sec:disc_conc}
	
	With the advent of large-scale optical spectroscopic surveys, a growing number of AGN have shown a change of their optical classification when observed years apart, driven by the appearance/disappearance of broad emission-line components \citep[e.g.,][]{Panda-Sniegowska2024}. More rarely, these objects have also been observed in markedly different optical/UV continuum spectral states. Such variability on time scales of a few years may be readily associated with changes in line-of-sight obscuration, while only in a minority of cases it has been safely linked to variations in the accretion regime \citep{Runnoe2016,MacLeod2016}. Nonetheless, the emerging `changing-state' picture \citep{RicciTrak2023} still presents several challenges. In particular, there is no well-established theoretical framework describing the structure and evolution of the variable, optically thick accretion disk, and observational constraints remain limited, especially on the far-UV and X-ray side. Moreover, existing studies are typically based on heterogeneous data sets or on sparse, non-uniform observations, which prevent an accurate tracking of the spectral evolution of individual sources across state transitions. As a consequence, the temporal connection between changes in the disk emission and the behavior of the X-ray corona is still poorly constrained.
	
	Here we have presented the results from a six-year (2019--2025) {\it Swift} monitoring campaign of the nearby type-1 Seyfert galaxy ESO\,511-G030, during which the observed flux of this source increased systematically by almost a factor of $\sim$\,10 in both the UV and X-rays (Fig.~\ref{figlcs}). A similar change is also observed spectroscopically, with a clear emergence of the accretion-disk emission component in the UV since 2021 (Fig.~\ref{fig:vfv}). 
	By considering the 2021 annual average of the AGN luminosity at 2500 \AA, which is close to the break observed in the $L_{\rm X}$--$L_{\rm UV}$ relation (Fig.~\ref{fig:LxLuv}, and assuming a bolometric correction of $\sim$\,5 after \citealt{Richards2006}, we estimate an Eddington ratio of just below 0.01 at the onset of the state transition. We stress that this is to be intended as an order-of-magnitude exercise, as the realistic uncertainty on the Eddington ratio is hardly smaller than a factor of $\sim$\,3--5, given the combined systematics on BH mass and bolometric correction. Even so, the resulting figure is in excellent agreement with the value that has been suggested to represent a universal threshold across which the accretion flow undergoes significant structural changes, similar to those observed in XRBs.
	
	Tracking this kind of transitions in AGN is extremely challenging, not just because of the expansion in the characteristic time scales. Indeed, in AGN only the low-temperature emission from relatively large radii in the disk is directly accessible in the optical/UV band. In the low $\lambda_{\mathrm{Edd}}$ regime, this emission can be easily confused with host-galaxy starlight and further dimmed and reddened by dust. X-ray obscuration typically represents another major source of uncertainty. Absorption and reddening effects are negligible in a bare object like ESO\,511-G030, which therefore stands out as a genuine, newly discovered changing-state AGN. A similar combination of extensive broadband monitoring and source evolution is almost unprecedented, comparable to what has been obtained in the last decade only for Mrk\,590 \citep[e.g.,][]{Palit2025,Palit2026}.
	
	The behavior of ESO\,511-G030 revealed by the ongoing {\it Swift} campaign is suggestive of a recovery of the inner accretion disk since 2019 after its apparent depletion in the previous decade. {\it Suzaku} observations in 2012 found the source in a persistently high X-ray state \citep{Ghosh2021}. Although no simultaneous UV data were available, most likely also the UV emission was still intense in that epoch. The lack of coverage between 2012 and 2019 prevents us from understanding how the source reached its lowest flux state ever recorded, but sets a stringent upper limit on the duration of the fading phase. Noting that most of the recovery took place in less than three years (2021--2023), it is not implausible that an even fainter state was reached before the start of this monitoring campaign. Interestingly, before the surge of the disk emission, the $L_{\rm X}$--$L_{\rm UV}$ relation in ESO\,511-G030 shows a flat or slightly negative slope, indicating a departure from the standard disk/corona interplay in two-phase models \citep[][]{Haardt1991,Haardt1993}. This trend is consistent with the inversion in the correlation between the UV–to–X-ray energy index $\alpha_{\rm ox}$ and $\lambda_{\rm Edd}$ found in population studies of changing-state AGN \citep[][]{Ruan2019}. The same applies to the relation between the X-ray photon index $\Gamma$ and $\lambda_{\rm Edd}$, which shows a clear convex shape, the most impressive example being that of NGC\,2617 \citep{Jana2026}. These slope changes are generally interpreted as indicative of a state transition, consisting in the replacement of the inner regions of an SS73-like disk with a hot, radiatively inefficient flow below a critical accretion rate. Within the systematic uncertainties, the break is always consistent with $\lambda_{\rm Edd}$\,$\sim$\,0.01, as also suggested by the evolution of the $L_{\rm X}$--$L_{\rm UV}$ relation in ESO\,511-G030. In this light, 
	the delay between the rise of the X-ray emission with respect to the UV one seen in 2021 is compatible with the picture of a progressive recovery of a standard X-ray corona following that of the inner accretion disk.

	Our {\it Swift} monitoring campaign of ESO\,511-G030 provides unambiguous evidence that the accretion-state transition observed in stellar-mass BH systems can occur in AGN likewise, and at a similar $\lambda_{\rm Edd}$. However, the strikingly short time scales of the declining ($<$\,7 yr) and rising ($\sim$\,2--3 yr) phases confirm that standard disk models return an imperfect description of the actual structure and workings of the optically thick component of the accretion flow in AGN. Various kinds of instabilities affecting the radiation-pressure dominated, innermost regions of the accretion disk have been invoked to explain the time scales of changing-state events in AGN \citep[e.g.,][]{Stern2018,Noda2018,Sniegowska2020}, yet any new generation of theoretical models must be informed by a regular monitoring during the state transitions. In a landscape where optical wide-field survey facilities like the Vera C.~Rubin Observatory \citep[][]{Ivezic2019} are expected to rapidly revolutionize the demography of changing-state AGN, the present campaign on ESO\,511-G030 demonstrates that simultaneous coverage up to the X-rays is essential to achieve a deeper understanding of this phenomenon.

	{\it \textbf{Acknowledgments:}} We thank the anonymous referee for their thorough reading of the manuscript and comments. RM acknowledges financial support from the INAF Scientific Directorate. This work has been partially supported by the ASI-INAF program I/004/11/6.
	
	
	\facilities{Neil Gehrels Swift Observatory}
	
	\appendix
	\section{Host-galaxy and AGN spectral decomposition}
	\label{floyds}
	
	Low-resolution spectroscopic observations of ESO\,511-G030 were obtained with the 2-m Faulkes Telescope North (FTN; Haleakala, Hawaii) and the identical 2-m Faulkes Telescope South (FTS; Siding Spring, Australia) in 2019, 2022, 2023, 2024, and 2025. The data were acquired with the FLOYDS spectrographs using a six-arcsec slit at the parallactic angle, and the resulting spectra were extracted using the AGN FLOYDS pipeline.\footnote{\url{https://github.com/svalenti/FLOYDS_pipeline}} To establish the absolute flux calibration, correct for telluric absorption features, and reduce fringing artifacts, we used the spectrum of the closest spectro-photometric standard star observed within 5 days. The FLOYDS spectrographs record first (red, $\sim$\,4000--10,000 \AA) and second (blue, $\sim$\,3500--5500 \AA) order spectra on the same CCD, and two exposures were typically requested per visit, ideally yielding four spectra (two blue\,+\,red pairs). However, some visits lacked a complete set of four spectra. Here we considered a total of 112 spectra: 31 and 28 for the blue and red FTN, respectively, and 26 and 27 for the blue and red FTS, respectively. Since residual fringing artifacts become noticeable redward of H$\alpha$, we clipped the spectra at a rest-frame wavelength of 7500 \AA. Moreover, an instrumental artifact is present around 4500 \AA, caused by a dip in sensitivity near the edge of the red arm of the spectrograph; this region was also excluded from the calibrated spectra. 
	Since the blue arm has twice the spectral resolution of the red arm (better resolving lines like [O\,\textsc{iii}]), and flux inter-calibration problems sometimes cause a mismatch in the overlapping spectral region, we fitted all the 112 spectra separately assuming Gaussian line profiles, a power-law AGN continuum with fixed slope of $\alpha_\lambda$\,=\,$-1.7$ (cf. Fig.~20 in \citealt{Calderone2017}), and a SWIRE Sc galaxy template, averaging the extrapolated host-galaxy luminosities at 2500 \AA\ obtained from each arm. 
	
	We considered the host-galaxy luminosities from the red-arm spectra to be more reliable, as this arm probes longer wavelengths where the host emission dominates. The host-galaxy contribution at rest-frame 2500 \AA\ ($L_{\rm UV,gal}$) of 2.5$\pm$0.3\,$\times$\,10$^{27}$ erg s$^{-1}$ Hz$^{-1}$ assumed in this work is therefore the one averaged over the red arms of FTN and FTS. The corresponding quantity derived from the blue arms is slightly lower, but consistent within the uncertainties.

	\bibliography{sample}

\begin{thebibliography}{}
\expandafter\ifx\csname natexlab\endcsname\relax\def\natexlab#1{#1}\fi
\providecommand{\url}[1]{\href{#1}{#1}}
\providecommand{\dodoi}[1]{doi:~\href{http://doi.org/#1}{\nolinkurl{#1}}}
\providecommand{\doeprint}[1]{\href{http://ascl.net/#1}{\nolinkurl{http://ascl.net/#1}}}
\providecommand{\doarXiv}[1]{\href{https://arxiv.org/abs/#1}{\nolinkurl{https://arxiv.org/abs/#1}}}

\bibitem[{M.~A. {Abramowicz} {et~al.}(1988){Abramowicz}, {Czerny}, {Lasota}, \&
  {Szuszkiewicz}}]{Abramowicz1988}
{Abramowicz}, M.~A., {Czerny}, B., {Lasota}, J.~P., \& {Szuszkiewicz}, E. 1988,
  \bibinfo{title}{{Slim Accretion Disks},} \apj, 332, 646,
  \dodoi{10.1086/166683}

\bibitem[{K.~A. {Arnaud}(1996){Arnaud}}]{Arnaud1996}
{Arnaud}, K.~A. 1996, \bibinfo{title}{{XSPEC: The First Ten Years},} in
  Astronomical Society of the Pacific Conference Series, Vol. 101, Astronomical
  Data Analysis Software and Systems V, ed. G.~H. {Jacoby} \& J.~{Barnes}, 17

\bibitem[{R.~C. {Bohlin} {et~al.}(1978){Bohlin}, {Savage}, \&
  {Drake}}]{Bohlin1978}
{Bohlin}, R.~C., {Savage}, B.~D., \& {Drake}, J.~F. 1978, \bibinfo{title}{{A
  survey of interstellar H I from Lalpha absorption measurements. II.},} \apj,
  224, 132, \dodoi{10.1086/156357}

\bibitem[{A.~Y.~K. {Bouquin} {et~al.}(2018){Bouquin}, {Gil de Paz},
  {Mu{\~n}oz-Mateos}, {Boissier}, {Sheth}, {Zaritsky}, {Peletier}, {Knapen}, \&
  {Gallego}}]{Bouquin2018}
{Bouquin}, A. Y.~K., {Gil de Paz}, A., {Mu{\~n}oz-Mateos}, J.~C., {et~al.}
  2018, \bibinfo{title}{{The GALEX/S$^{4}$G Surface Brightness and Color
  Profiles Catalog. I. Surface Photometry and Color Gradients of Galaxies},}
  \apjs, 234, 18, \dodoi{10.3847/1538-4365/aaa384}

\bibitem[{A.~A. {Breeveld} {et~al.}(2011){Breeveld}, {Landsman}, {Holland},
  {Roming}, {Kuin}, \& {Page}}]{Breeveld2011}
{Breeveld}, A.~A., {Landsman}, W., {Holland}, S.~T., {et~al.} 2011,
  \bibinfo{title}{{An Updated Ultraviolet Calibration for the Swift/UVOT},} in
  American Institute of Physics Conference Series, Vol. 1358, Gamma Ray Bursts
  2010, ed. J.~E. {McEnery}, J.~L. {Racusin}, \& N.~{Gehrels} (AIP), 373--376,
  \dodoi{10.1063/1.3621807}

\bibitem[{D.~N. {Burrows} {et~al.}(2005){Burrows}, {Hill}, {Nousek}, {Kennea},
  {Wells}, {Osborne}, {Abbey}, {Beardmore}, {Mukerjee}, {Short}, {Chincarini},
  {Campana}, {Citterio}, {Moretti}, {Pagani}, {Tagliaferri}, {Giommi},
  {Capalbi}, {Tamburelli}, {Angelini}, {Cusumano}, {Br{\"a}uninger}, {Burkert},
  \& {Hartner}}]{Burrows2005}
{Burrows}, D.~N., {Hill}, J.~E., {Nousek}, J.~A., {et~al.} 2005,
  \bibinfo{title}{{The Swift X-Ray Telescope},} \ssr, 120, 165,
  \dodoi{10.1007/s11214-005-5097-2}

\bibitem[{Z.-Y. {Cai} \& J.-X. {Wang}(2023){Cai} \& {Wang}}]{CaiWang2023}
{Cai}, Z.-Y., \& {Wang}, J.-X. 2023, \bibinfo{title}{{A universal average
  spectral energy distribution for quasars from the optical to the extreme
  ultraviolet},} Nature Astronomy, 7, 1506, \dodoi{10.1038/s41550-023-02088-5}

\bibitem[{G. {Calderone} {et~al.}(2017){Calderone}, {Nicastro}, {Ghisellini},
  {Dotti}, {Sbarrato}, {Shankar}, \& {Colpi}}]{Calderone2017}
{Calderone}, G., {Nicastro}, L., {Ghisellini}, G., {et~al.} 2017,
  \bibinfo{title}{{QSFIT: automatic analysis of optical AGN spectra},} \mnras,
  472, 4051, \dodoi{10.1093/mnras/stx2239}

\bibitem[{G. {de Vaucouleurs} {et~al.}(1991){de Vaucouleurs}, {de Vaucouleurs},
  {Corwin}, {Buta}, {Paturel}, \& {Fouque}}]{deVaucouleurs1991}
{de Vaucouleurs}, G., {de Vaucouleurs}, A., {Corwin}, Jr., H.~G., {et~al.}
  1991, {Third Reference Catalogue of Bright Galaxies}

\bibitem[{C. {Done} {et~al.}(2007){Done}, {Gierli{\'n}ski}, \&
  {Kubota}}]{Done2007}
{Done}, C., {Gierli{\'n}ski}, M., \& {Kubota}, A. 2007,
  \bibinfo{title}{{Modelling the behaviour of accretion flows in X-ray
  binaries. Everything you always wanted to know about accretion but were
  afraid to ask},} \aapr, 15, 1, \dodoi{10.1007/s00159-007-0006-1}

\bibitem[{D. {Foreman-Mackey} {et~al.}(2013){Foreman-Mackey}, {Hogg}, {Lang},
  \& {Goodman}}]{Foreman-Mackey2013}
{Foreman-Mackey}, D., {Hogg}, D.~W., {Lang}, D., \& {Goodman}, J. 2013,
  \bibinfo{title}{{emcee: The MCMC Hammer},} \pasp, 125, 306,
  \dodoi{10.1086/670067}

\bibitem[{N. {Gehrels} {et~al.}(2004){Gehrels}, {Chincarini}, {Giommi},
  {Mason}, {Nousek}, {Wells}, {White}, {Barthelmy}, {Burrows}, {Cominsky},
  {Hurley}, {Marshall}, {M{\'e}sz{\'a}ros}, {Roming}, {Angelini}, {Barbier},
  {Belloni}, {Campana}, {Caraveo}, {Chester}, {Citterio}, {Cline}, {Cropper},
  {Cummings}, {Dean}, {Feigelson}, {Fenimore}, {Frail}, {Fruchter}, {Garmire},
  {Gendreau}, {Ghisellini}, {Greiner}, {Hill}, {Hunsberger}, {Krimm},
  {Kulkarni}, {Kumar}, {Lebrun}, {Lloyd-Ronning}, {Markwardt}, {Mattson},
  {Mushotzky}, {Norris}, {Osborne}, {Paczynski}, {Palmer}, {Park}, {Parsons},
  {Paul}, {Rees}, {Reynolds}, {Rhoads}, {Sasseen}, {Schaefer}, {Short},
  {Smale}, {Smith}, {Stella}, {Tagliaferri}, {Takahashi}, {Tashiro},
  {Townsley}, {Tueller}, {Turner}, {Vietri}, {Voges}, {Ward}, {Willingale},
  {Zerbi}, \& {Zhang}}]{Gehrels2004}
{Gehrels}, N., {Chincarini}, G., {Giommi}, P., {et~al.} 2004,
  \bibinfo{title}{{The Swift Gamma-Ray Burst Mission},} \apj, 611, 1005,
  \dodoi{10.1086/422091}

\bibitem[{R. {Ghosh} \& S. {Laha}(2021){Ghosh} \& {Laha}}]{Ghosh2021}
{Ghosh}, R., \& {Laha}, S. 2021, \bibinfo{title}{{An X-Ray Spectral Study of
  the Origin of Reflection Features in Bare Seyfert 1 Galaxy ESO 511-G030},}
  \apj, 908, 198, \dodoi{10.3847/1538-4357/abd40c}

\bibitem[{F. {Haardt} \& L. {Maraschi}(1991){Haardt} \&
  {Maraschi}}]{Haardt1991}
{Haardt}, F., \& {Maraschi}, L. 1991, \bibinfo{title}{{A Two-Phase Model for
  the X-Ray Emission from Seyfert Galaxies},} \apjl, 380, L51,
  \dodoi{10.1086/186171}

\bibitem[{F. {Haardt} \& L. {Maraschi}(1993){Haardt} \&
  {Maraschi}}]{Haardt1993}
{Haardt}, F., \& {Maraschi}, L. 1993, \bibinfo{title}{{X-Ray Spectra from
  Two-Phase Accretion Disks},} \apj, 413, 507, \dodoi{10.1086/173020}

\bibitem[{S. {Hagen} {et~al.}(2024){Hagen}, {Done}, {Silverman}, {Li}, {Liu},
  {Ren}, {Buchner}, {Merloni}, {Nagao}, \& {Salvato}}]{Hagen2024}
{Hagen}, S., {Done}, C., {Silverman}, J.~D., {et~al.} 2024,
  \bibinfo{title}{{Systematic collapse of the accretion disc across the
  supermassive black hole population},} \mnras, 534, 2803,
  \dodoi{10.1093/mnras/stae2272}

\bibitem[{ {HI4PI Collaboration} {et~al.}(2016){HI4PI Collaboration}, {Ben
  Bekhti}, {Fl{\"o}er}, {Keller}, {Kerp}, {Lenz}, {Winkel}, {Bailin},
  {Calabretta}, {Dedes}, {Ford}, {Gibson}, {Haud}, {Janowiecki}, {Kalberla},
  {Lockman}, {McClure-Griffiths}, {Murphy}, {Nakanishi}, {Pisano}, \&
  {Staveley-Smith}}]{HI4PI2016}
{HI4PI Collaboration}, {Ben Bekhti}, N., {Fl{\"o}er}, L., {et~al.} 2016,
  \bibinfo{title}{{HI4PI: A full-sky H I survey based on EBHIS and GASS},}
  \aap, 594, A116, \dodoi{10.1051/0004-6361/201629178}

\bibitem[{P.~F. {Hopkins} {et~al.}(2004){Hopkins}, {Strauss}, {Hall},
  {Richards}, {Cooper}, {Schneider}, {Vanden Berk}, {Jester}, {Brinkmann}, \&
  {Szokoly}}]{Hopkins2004}
{Hopkins}, P.~F., {Strauss}, M.~A., {Hall}, P.~B., {et~al.} 2004,
  \bibinfo{title}{{Dust Reddening in Sloan Digital Sky Survey Quasars},} \aj,
  128, 1112, \dodoi{10.1086/423291}

\bibitem[{{\v{Z}}. {Ivezi{\'c}} {et~al.}(2019){Ivezi{\'c}}, {Kahn}, {Tyson},
  {Abel}, {Acosta}, {Allsman}, {Alonso}, {AlSayyad}, {Anderson}, {Andrew},
  {Angel}, {Angeli}, {Ansari}, {Antilogus}, {Araujo}, {Armstrong}, {Arndt},
  {Astier}, {Aubourg}, {Auza}, {Axelrod}, {Bard}, {Barr}, {Barrau}, {Bartlett},
  {Bauer}, {Bauman}, {Baumont}, {Bechtol}, {Bechtol}, {Becker}, {Becla},
  {Beldica}, {Bellavia}, {Bianco}, {Biswas}, {Blanc}, {Blazek}, {Blandford},
  {Bloom}, {Bogart}, {Bond}, {Booth}, {Borgland}, {Borne}, {Bosch}, {Boutigny},
  {Brackett}, {Bradshaw}, {Brandt}, {Brown}, {Bullock}, {Burchat}, {Burke},
  {Cagnoli}, {Calabrese}, {Callahan}, {Callen}, {Carlin}, {Carlson},
  {Chandrasekharan}, {Charles-Emerson}, {Chesley}, {Cheu}, {Chiang}, {Chiang},
  {Chirino}, {Chow}, {Ciardi}, {Claver}, {Cohen-Tanugi}, {Cockrum}, {Coles},
  {Connolly}, {Cook}, {Cooray}, {Covey}, {Cribbs}, {Cui}, {Cutri}, {Daly},
  {Daniel}, {Daruich}, {Daubard}, {Daues}, {Dawson}, {Delgado}, {Dellapenna},
  {de Peyster}, {de Val-Borro}, {Digel}, {Doherty}, {Dubois},
  {Dubois-Felsmann}, {Durech}, {Economou}, {Eifler}, {Eracleous}, {Emmons},
  {Fausti Neto}, {Ferguson}, {Figueroa}, {Fisher-Levine}, {Focke}, {Foss},
  {Frank}, {Freemon}, {Gangler}, {Gawiser}, {Geary}, {Gee}, {Geha}, {Gessner},
  {Gibson}, {Gilmore}, {Glanzman}, {Glick}, {Goldina}, {Goldstein}, {Goodenow},
  {Graham}, {Gressler}, {Gris}, {Guy}, {Guyonnet}, {Haller}, {Harris},
  {Hascall}, {Haupt}, {Hernandez}, {Herrmann}, {Hileman}, {Hoblitt}, {Hodgson},
  {Hogan}, {Howard}, {Huang}, {Huffer}, {Ingraham}, {Innes}, {Jacoby}, {Jain},
  {Jammes}, {Jee}, {Jenness}, {Jernigan}, {Jevremovi{\'c}}, {Johns}, {Johnson},
  {Johnson}, {Jones}, {Juramy-Gilles}, {Juri{\'c}}, {Kalirai}, {Kallivayalil},
  {Kalmbach}, {Kantor}, {Karst}, {Kasliwal}, {Kelly}, {Kessler}, {Kinnison},
  {Kirkby}, {Knox}, {Kotov}, {Krabbendam}, {Krughoff}, {Kub{\'a}nek},
  {Kuczewski}, {Kulkarni}, {Ku}, {Kurita}, {Lage}, {Lambert}, {Lange},
  {Langton}, {Le Guillou}, {Levine}, {Liang}, {Lim}, {Lintott}, {Long},
  {Lopez}, {Lotz}, {Lupton}, {Lust}, {MacArthur}, {Mahabal}, {Mandelbaum},
  {Markiewicz}, {Marsh}, {Marshall}, {Marshall}, {May}, {McKercher}, {McQueen},
  {Meyers}, {Migliore}, {Miller}, \& {Mills}}]{Ivezic2019}
{Ivezi{\'c}}, {\v{Z}}., {Kahn}, S.~M., {Tyson}, J.~A., {et~al.} 2019,
  \bibinfo{title}{{LSST: From Science Drivers to Reference Design and
  Anticipated Data Products},} \apj, 873, 111, \dodoi{10.3847/1538-4357/ab042c}

\bibitem[{A. {Jana} {et~al.}(2026){Jana}, {Ricci}, {Tortosa}, {Dimopoulos},
  {Trakhtenbrot}, {Bauer}, {Temple}, {Koss}, {Gupta}, {Chang}, {Diaz}, {Illic},
  {Kallov{\'a}}, \& {Shablovinskaya}}]{Jana2026}
{Jana}, A., {Ricci}, C., {Tortosa}, A., {et~al.} 2026,
  \bibinfo{title}{{Multiwavelength properties of changing-state active galactic
  nuclei: I. The evolution of soft excess and X-ray continuum},} \aap, 707,
  A213, \dodoi{10.1051/0004-6361/202556654}

\bibitem[{J.-L. {Kang} {et~al.}(2025){Kang}, {Done}, {Hagen}, {Temple},
  {Silverman}, {Li}, \& {Liu}}]{Kang2025}
{Kang}, J.-L., {Done}, C., {Hagen}, S., {et~al.} 2025,
  \bibinfo{title}{{Systematic collapse of the accretion disc in AGN confirmed
  by UV photometry and broad-line spectra},} \mnras, 538, 121,
  \dodoi{10.1093/mnras/staf145}

\bibitem[{R.~E. {Kass} \& A.~E. {Raftery}(1995){Kass} \& {Raftery}}]{Kass1995}
{Kass}, R.~E., \& {Raftery}, A.~E. 1995, \bibinfo{title}{{Bayes Factors},}
  Journal of the American Statistical Association, 90, 773,
  \dodoi{10.1080/01621459.1995.10476572}

\bibitem[{M.~J. {Koss} {et~al.}(2022){Koss}, {Ricci}, {Trakhtenbrot}, {Oh},
  {den Brok}, {Mej{\'\i}a-Restrepo}, {Stern}, {Privon}, {Treister}, {Powell},
  {Mushotzky}, {Bauer}, {Ananna}, {Balokovi{\'c}}, {B{\"a}r}, {Becker},
  {Bessiere}, {Burtscher}, {Caglar}, {Congiu}, {Evans}, {Harrison}, {Heida},
  {Ichikawa}, {Kamraj}, {Lamperti}, {Pacucci}, {Ricci}, {Riffel}, {Rojas},
  {Schawinski}, {Temple}, {Urry}, {Veilleux}, \& {Williams}}]{Koss2022}
{Koss}, M.~J., {Ricci}, C., {Trakhtenbrot}, B., {et~al.} 2022,
  \bibinfo{title}{{BASS. XXII. The BASS DR2 AGN Catalog and Data},} \apjs, 261,
  2, \dodoi{10.3847/1538-4365/ac6c05}

\bibitem[{S. {Laha} {et~al.}(2014){Laha}, {Guainazzi}, {Dewangan},
  {Chakravorty}, \& {Kembhavi}}]{Laha2014}
{Laha}, S., {Guainazzi}, M., {Dewangan}, G.~C., {Chakravorty}, S., \&
  {Kembhavi}, A.~K. 2014, \bibinfo{title}{{Warm absorbers in X-rays (WAX), a
  comprehensive high-resolution grating spectral study of a sample of Seyfert
  galaxies - I. A global view and frequency of occurrence of warm absorbers.},}
  \mnras, 441, 2613, \dodoi{10.1093/mnras/stu669}

\bibitem[{A. {Lauberts}(1982){Lauberts}}]{Lauberts1982}
{Lauberts}, A. 1982, {ESO/Uppsala survey of the ESO(B) atlas}

\bibitem[{E. {Lusso} \& G. {Risaliti}(2016){Lusso} \& {Risaliti}}]{Lusso2016}
{Lusso}, E., \& {Risaliti}, G. 2016, \bibinfo{title}{{The Tight Relation
  between X-Ray and Ultraviolet Luminosity of Quasars},} \apj, 819, 154,
  \dodoi{10.3847/0004-637X/819/2/154}

\bibitem[{E. {Lusso} {et~al.}(2020){Lusso}, {Risaliti}, {Nardini},
  {Bargiacchi}, {Benetti}, {Bisogni}, {Capozziello}, {Civano}, {Eggleston},
  {Elvis}, {Fabbiano}, {Gilli}, {Marconi}, {Paolillo}, {Piedipalumbo},
  {Salvestrini}, {Signorini}, \& {Vignali}}]{Lusso2020}
{Lusso}, E., {Risaliti}, G., {Nardini}, E., {et~al.} 2020,
  \bibinfo{title}{{Quasars as standard candles. III. Validation of a new sample
  for cosmological studies},} \aap, 642, A150,
  \dodoi{10.1051/0004-6361/202038899}

\bibitem[{C.~L. {MacLeod} {et~al.}(2016){MacLeod}, {Ross}, {Lawrence}, {Goad},
  {Horne}, {Burgett}, {Chambers}, {Flewelling}, {Hodapp}, {Kaiser}, {Magnier},
  {Wainscoat}, \& {Waters}}]{MacLeod2016}
{MacLeod}, C.~L., {Ross}, N.~P., {Lawrence}, A., {et~al.} 2016,
  \bibinfo{title}{{A systematic search for changing-look quasars in SDSS},}
  \mnras, 457, 389, \dodoi{10.1093/mnras/stv2997}

\bibitem[{R. {Maiolino} {et~al.}(2001){Maiolino}, {Marconi}, {Salvati},
  {Risaliti}, {Severgnini}, {Oliva}, {La Franca}, \& {Vanzi}}]{Maiolino2001}
{Maiolino}, R., {Marconi}, A., {Salvati}, M., {et~al.} 2001,
  \bibinfo{title}{{Dust in active nuclei. I. Evidence for ``anomalous''
  properties},} \aap, 365, 28, \dodoi{10.1051/0004-6361:20000177}

\bibitem[{R. {Middei} {et~al.}(2023){Middei}, {Petrucci}, {Bianchi}, {Ursini},
  {Matzeu}, {Vagnetti}, {Tortosa}, {Marinucci}, {Matt}, {Piconcelli}, {De
  Rosa}, {De Marco}, {Reeves}, {Perri}, {Guainazzi}, {Cappi}, \&
  {Done}}]{Middei2023}
{Middei}, R., {Petrucci}, P.-O., {Bianchi}, S., {et~al.} 2023,
  \bibinfo{title}{{Tracking the spectral properties of ESO 511-G030 across
  different epochs},} \aap, 672, A101, \dodoi{10.1051/0004-6361/202244022}

\bibitem[{R. {Narayan} \& I. {Yi}(1995){Narayan} \& {Yi}}]{Narayan1995}
{Narayan}, R., \& {Yi}, I. 1995, \bibinfo{title}{{Advection-dominated
  Accretion: Underfed Black Holes and Neutron Stars},} \apj, 452, 710,
  \dodoi{10.1086/176343}

\bibitem[{H. {Noda} \& C. {Done}(2018){Noda} \& {Done}}]{Noda2018}
{Noda}, H., \& {Done}, C. 2018, \bibinfo{title}{{Explaining changing-look AGN
  with state transition triggered by rapid mass accretion rate drop},} \mnras,
  480, 3898, \dodoi{10.1093/mnras/sty2032}

\bibitem[{B. {Palit} {et~al.}(2025){Palit}, {{\'S}niegowska}, {Markowitz},
  {R{\'o}{\.z}a{\'n}ska}, {Farah}, \& {Howell}}]{Palit2025}
{Palit}, B., {{\'S}niegowska}, M., {Markowitz}, A., {et~al.} 2025,
  \bibinfo{title}{{Markarian 590: the AGN awakens},} \mnras, 540, L14,
  \dodoi{10.1093/mnrasl/slaf027}

\bibitem[{B. {Palit} {et~al.}(2026){Palit}, {Rozanska}, {Markowitz}, {Lawther},
  {Vestergaard}, {Ruan}, {Saha}, {Walsh}, {Borkar}, {Sniegowska}, \&
  {Lu}}]{Palit2026}
{Palit}, B., {Rozanska}, A., {Markowitz}, A.~G., {et~al.} 2026,
  \bibinfo{title}{{The spectral state transition of Mkn 590, a potential link
  between AGNs and X-ray binaries?},} arXiv e-prints, arXiv:2602.17572,
  \dodoi{10.48550/arXiv.2602.17572}

\bibitem[{S. {Panda} \& M. {{\'S}niegowska}(2024){Panda} \&
  {{\'S}niegowska}}]{Panda-Sniegowska2024}
{Panda}, S., \& {{\'S}niegowska}, M. 2024, \bibinfo{title}{{Changing-look
  Active Galactic Nuclei. I. Tracking the Transition on the Main Sequence of
  Quasars},} \apjs, 272, 13, \dodoi{10.3847/1538-4365/ad344f}

\bibitem[{M. {Polletta} {et~al.}(2007){Polletta}, {Tajer}, {Maraschi},
  {Trinchieri}, {Lonsdale}, {Chiappetti}, {Andreon}, {Pierre}, {Le F{\`e}vre},
  {Zamorani}, {Maccagni}, {Garcet}, {Surdej}, {Franceschini}, {Alloin},
  {Shupe}, {Surace}, {Fang}, {Rowan-Robinson}, {Smith}, \&
  {Tresse}}]{Polletta2007}
{Polletta}, M., {Tajer}, M., {Maraschi}, L., {et~al.} 2007,
  \bibinfo{title}{{Spectral Energy Distributions of Hard X-Ray Selected Active
  Galactic Nuclei in the XMM-Newton Medium Deep Survey},} \apj, 663, 81,
  \dodoi{10.1086/518113}

\bibitem[{J. {Poutanen} {et~al.}(2007){Poutanen}, {Lipunova}, {Fabrika},
  {Butkevich}, \& {Abolmasov}}]{Poutanen2007}
{Poutanen}, J., {Lipunova}, G., {Fabrika}, S., {Butkevich}, A.~G., \&
  {Abolmasov}, P. 2007, \bibinfo{title}{{Supercritically accreting stellar mass
  black holes as ultraluminous X-ray sources},} \mnras, 377, 1187,
  \dodoi{10.1111/j.1365-2966.2007.11668.x}

\bibitem[{M.~L. {Prevot} {et~al.}(1984){Prevot}, {Lequeux}, {Maurice},
  {Prevot}, \& {Rocca-Volmerange}}]{Prevot1984}
{Prevot}, M.~L., {Lequeux}, J., {Maurice}, E., {Prevot}, L., \&
  {Rocca-Volmerange}, B. 1984, \bibinfo{title}{{The typical interstellar
  extinction in the Small Magellanic Cloud.},} \aap, 132, 389

\bibitem[{C. {Ricci} \& B. {Trakhtenbrot}(2023){Ricci} \&
  {Trakhtenbrot}}]{RicciTrak2023}
{Ricci}, C., \& {Trakhtenbrot}, B. 2023, \bibinfo{title}{{Changing-look active
  galactic nuclei},} Nature Astronomy, 7, 1282,
  \dodoi{10.1038/s41550-023-02108-4}

\bibitem[{G.~T. {Richards} {et~al.}(2006){Richards}, {Lacy},
  {Storrie-Lombardi}, {Hall}, {Gallagher}, {Hines}, {Fan}, {Papovich}, {Vanden
  Berk}, {Trammell}, {Schneider}, {Vestergaard}, {York}, {Jester}, {Anderson},
  {Budav{\'a}ri}, \& {Szalay}}]{Richards2006}
{Richards}, G.~T., {Lacy}, M., {Storrie-Lombardi}, L.~J., {et~al.} 2006,
  \bibinfo{title}{{Spectral Energy Distributions and Multiwavelength Selection
  of Type 1 Quasars},} \apjs, 166, 470, \dodoi{10.1086/506525}

\bibitem[{P.~W.~A. {Roming} {et~al.}(2005){Roming}, {Kennedy}, {Mason},
  {Nousek}, {Ahr}, {Bingham}, {Broos}, {Carter}, {Hancock}, {Huckle},
  {Hunsberger}, {Kawakami}, {Killough}, {Koch}, {McLelland}, {Smith}, {Smith},
  {Soto}, {Boyd}, {Breeveld}, {Holland}, {Ivanushkina}, {Pryzby}, {Still}, \&
  {Stock}}]{Roming2005}
{Roming}, P. W.~A., {Kennedy}, T.~E., {Mason}, K.~O., {et~al.} 2005,
  \bibinfo{title}{{The Swift Ultra-Violet/Optical Telescope},} \ssr, 120, 95,
  \dodoi{10.1007/s11214-005-5095-4}

\bibitem[{J.~J. {Ruan} {et~al.}(2019){Ruan}, {Anderson}, {Eracleous}, {Green},
  {Haggard}, {MacLeod}, {Runnoe}, \& {Sobolewska}}]{Ruan2019}
{Ruan}, J.~J., {Anderson}, S.~F., {Eracleous}, M., {et~al.} 2019,
  \bibinfo{title}{{The Analogous Structure of Accretion Flows in Supermassive
  and Stellar Mass Black Holes: New Insights from Faded Changing-look
  Quasars},} \apj, 883, 76, \dodoi{10.3847/1538-4357/ab3c1a}

\bibitem[{J.~C. {Runnoe} {et~al.}(2016){Runnoe}, {Cales}, {Ruan}, {Eracleous},
  {Anderson}, {Shen}, {Green}, {Morganson}, {LaMassa}, {Greene}, {Dwelly},
  {Schneider}, {Merloni}, {Georgakakis}, \& {Roman-Lopes}}]{Runnoe2016}
{Runnoe}, J.~C., {Cales}, S., {Ruan}, J.~J., {et~al.} 2016,
  \bibinfo{title}{{Now you see it, now you don't: the disappearing central
  engine of the quasar J1011+5442},} \mnras, 455, 1691,
  \dodoi{10.1093/mnras/stv2385}

\bibitem[{I. {Saccheo} {et~al.}(2023){Saccheo}, {Bongiorno}, {Piconcelli},
  {Testa}, {Bischetti}, {Bisogni}, {Bruni}, {Cresci}, {Feruglio}, {Fiore},
  {Grazian}, {Luminari}, {Lusso}, {Mainieri}, {Maiolino}, {Marconi}, {Ricci},
  {Tombesi}, {Travascio}, {Vietri}, {Vignali}, {Zappacosta}, \& {La
  Franca}}]{Saccheo2023}
{Saccheo}, I., {Bongiorno}, A., {Piconcelli}, E., {et~al.} 2023,
  \bibinfo{title}{{The WISSH quasars project. XI. The mean spectral energy
  distribution and bolometric corrections of the most luminous quasars},} \aap,
  671, A34, \dodoi{10.1051/0004-6361/202244296}

\bibitem[{E.~F. {Schlafly} \& D.~P. {Finkbeiner}(2011){Schlafly} \&
  {Finkbeiner}}]{Schlafly2011}
{Schlafly}, E.~F., \& {Finkbeiner}, D.~P. 2011, \bibinfo{title}{{Measuring
  Reddening with Sloan Digital Sky Survey Stellar Spectra and Recalibrating
  SFD},} \apj, 737, 103, \dodoi{10.1088/0004-637X/737/2/103}

\bibitem[{N.~I. {Shakura} \& R.~A. {Sunyaev}(1973){Shakura} \&
  {Sunyaev}}]{Shakura1973}
{Shakura}, N.~I., \& {Sunyaev}, R.~A. 1973, \bibinfo{title}{{Black holes in
  binary systems. Observational appearance.},} \aap, 24, 337

\bibitem[{G.~A. {Shields}(1978){Shields}}]{Shields1978}
{Shields}, G.~A. 1978, \bibinfo{title}{{Thermal continuum from accretion disks
  in quasars},} \nat, 272, 706, \dodoi{10.1038/272706a0}

\bibitem[{M. {{\'S}niegowska} {et~al.}(2020){{\'S}niegowska}, {Czerny}, {Bon},
  \& {Bon}}]{Sniegowska2020}
{{\'S}niegowska}, M., {Czerny}, B., {Bon}, E., \& {Bon}, N. 2020,
  \bibinfo{title}{{Possible mechanism for multiple changing-look phenomena in
  active galactic nuclei},} \aap, 641, A167,
  \dodoi{10.1051/0004-6361/202038575}

\bibitem[{D. {Stern} {et~al.}(2018){Stern}, {McKernan}, {Graham}, {Ford},
  {Ross}, {Meisner}, {Assef}, {Balokovi{\'c}}, {Brightman}, {Dey}, {Drake},
  {Djorgovski}, {Eisenhardt}, \& {Jun}}]{Stern2018}
{Stern}, D., {McKernan}, B., {Graham}, M.~J., {et~al.} 2018, \bibinfo{title}{{A
  Mid-IR Selected Changing-look Quasar and Physical Scenarios for Abrupt AGN
  Fading},} \apj, 864, 27, \dodoi{10.3847/1538-4357/aac726}

\bibitem[{J. {Svoboda} {et~al.}(2017){Svoboda}, {Guainazzi}, \&
  {Merloni}}]{Svoboda2017}
{Svoboda}, J., {Guainazzi}, M., \& {Merloni}, A. 2017, \bibinfo{title}{{AGN
  spectral states from simultaneous UV and X-ray observations by XMM-Newton},}
  \aap, 603, A127, \dodoi{10.1051/0004-6361/201630181}

\bibitem[{B. {Trakhtenbrot} {et~al.}(2017){Trakhtenbrot}, {Ricci}, {Koss},
  {Schawinski}, {Mushotzky}, {Ueda}, {Veilleux}, {Lamperti}, {Oh}, {Treister},
  {Stern}, {Harrison}, {Balokovi{\'c}}, \& {Gehrels}}]{Trakhtenbrot2017}
{Trakhtenbrot}, B., {Ricci}, C., {Koss}, M.~J., {et~al.} 2017,
  \bibinfo{title}{{BAT AGN Spectroscopic Survey (BASS) - VI. The
  {\ensuremath{\Gamma}}$_{X}$-L/L$_{Edd}$ relation},} \mnras, 470, 800,
  \dodoi{10.1093/mnras/stx1117}

\bibitem[{J.-M. {Wang} \& Y.-Y. {Zhou}(1999){Wang} \& {Zhou}}]{Wang1999}
{Wang}, J.-M., \& {Zhou}, Y.-Y. 1999, \bibinfo{title}{{Self-similar Solution of
  Optically Thick Advection-dominated Flows},} \apj, 516, 420,
  \dodoi{10.1086/307080}

\bibitem[{S.~S. {Wilks}(1938){Wilks}}]{Wilks1938}
{Wilks}, S.~S. 1938, \bibinfo{title}{{The Large-Sample Distribution of the
  Likelihood Ratio for Testing Composite Hypotheses},} The Annals of
  Mathematical Statistics, 9, 60, \dodoi{10.1214/aoms/1177732360}

\end{thebibliography}
	\bibliographystyle{aasjournalv7}
\end{document}